\documentclass[pra,reprint,showpacs]{revtex4-1}
\usepackage[T1]{fontenc}
\usepackage[latin9]{inputenc}
\usepackage{color}
\usepackage{amsthm}
\usepackage{amsmath}
\usepackage{amssymb}
\usepackage{graphicx}

\theoremstyle{plain}
\newtheorem{thm}{\protect\theoremname}
\providecommand{\theoremname}{Theorem}

\begin{document}
\global\long\global\long\global\long\def\bra#1{\mbox{\ensuremath{\langle#1|}}}
\global\long\global\long\global\long\def\ket#1{\mbox{\ensuremath{|#1\rangle}}}
\global\long\global\long\global\long\def\bk#1#2{\mbox{\ensuremath{\ensuremath{\langle#1|#2\rangle}}}}
\global\long\global\long\global\long\def\kb#1#2{\mbox{\ensuremath{\ensuremath{\ensuremath{|#1\rangle\!\langle#2|}}}}}

\title{Fraction of isospectral states exhibiting quantum correlations}

\author{Micha\l{} Oszmaniec}

\email{oszmaniec@cft.edu.pl}

\author{Marek Ku\'{s}}

\address{Center for Theoretical Physics, Polish Academy of Sciences, Al. Lotników
32/46, 02-668 Warszawa}

\begin{abstract}
For several types of correlations: mixed-state entanglement in systems of
distinguishable particles, particle entanglement in systems of
indistinguishable bosons and fermions and non-Gaussian correlations in
fermionic systems we estimate the fraction of non-correlated states among the
density matrices with the same spectra. We prove that for the purity exceeding
some critical value (depending on the considered problem) fraction of
non-correlated states tends to zero exponentially fast with the dimension of
the relevant Hilbert space. As a consequence a state randomly chosen from the
set of density matrices possessing the same spectra is asymptotically a
correlated one. To prove this we developed a systematic framework for detection
of correlations via nonlinear witnesses.

\end{abstract}


\maketitle

The notion of quantum correlations in physical systems is a concept that
depends both on the system as well as on the physical property in question.
Taking as a paradigmatic example a familiar notion of entanglement in systems
of distinguishable particles \cite{01horo rev}, we may construct a general
scheme of defining quantum correlations. We start with a class of pure states
$\mathcal{M}$ lacking the desired correlation property (in the case of
entanglement these are all pure product states). The non-correlated
(non-entangled) mixed states are further defined as statistical mixtures of
pure states taken from the chosen class $\mathcal{M}$. All other states are
then called correlated (entangled).

The same scheme can be extended to other interesting cases by modifying the
choice of the class $\mathcal{M}$ of `non-correlated' pure states. For
indistinguishable particles the indispensable (anti)symmetrization of the
wave-function under permutation of subsystems introduces strong quantum
correlations.  Nevertheless one can pose a
legitimate question about nature of correlations which go beyond the mere fact
that the states are anti(symmetric). In fact, as recently shown in \cite{Activation of bosons}, in the case of bosons such
correlations  can be extracted into an entangled state of distinguishable
subsystems represented by independent modes. To analyze a role of double-occupancy
errors for operation of quantum gates composed of two quantum dots, the authors
of \cite{schliemann01} introduced a measure of correlations in fermionic
two-particle systems. It ascribes the vanishing entanglement (or, in other
words, vanishing correlations) only to pure states which are expressible in
terms of a single Slater determinant. The pure non-correlated states are again
probabilistic mixtures of non-correlated pure ones. This construction was
generalized in \cite{Eckert} to fermionic and bosonic systems of arbitrary
fixed number of particles occupying a finite number of one-particle states.
Here the underlying Hilbert space is no longer a product of Hilbert spaces of
individual subsystems, but rather an antisymmetric or symmetric part of it. The
class $\mathcal{M}$ of non-correlated pure states consists of, respectively,
the states in the form of a single Slater determinant and the product bosonic
states. 

Quantum information theory with bosons \cite{bos gaussian}, and fermions
\cite{ferm lagr,ferm gauss master,ferm noisy} in the Gaussian settings, where
the number of particles can vary, is another area where we can apply the above
scheme to discriminate non-correlated and correlated states. The underlying
Hilbert spaces are the bosonic and fermionic Fock spaces,  whereas the classes
of non-correlated pure states are obtained from the vacuum state by actions of
Hamiltonians quadratic in, respectively, bosonic and fermionic creation and
annihilation operators.  In the later case states that are not-correlated are
related to computation protocols performed with Majorana fermions that can be
classically simulated \cite{ferm noisy}. Moreover, states that cannot be
written as a statistical combination of fermionic Gaussian states are states
not described by the Bogolyubov mean field theory \cite{gauss nap}.

Although in the following we concentrate on correlations in the systems
mentioned above (distinguishable and non-distinguishable particles, fermionic
Gaussian states) it is worth mentioning that the outlined general scheme of
defining (non)-correlated states can be further extended to encompass, e.g.,
$k$-separabile states \cite{k sep}. Another important type of correlations that
can be analyzed within the same frame are `non-classical' properties of light
\cite{glauber}. Here the class $\mathcal{M}$ consist of Glauber coherent states
and `classical' (`non-correlated') mixed states are precisely those having a
positive P-representation. This notion of classicality was extended to spin
states \cite{giraud08}, where again the same construction applies
\cite{quasiclassical}.



In all considered cases the correlation properties are invariant with respect
to specific classes of transformations performed on the system in question.
Thus, e.g., entanglement of distinguishable particles is unchanged under local
unitary transformations. In all cases such correlation-preserving operations
form a proper subset of all (global) unitary transformation that can be applied
to the whole system. Global unitary transformation preserve the spectrum of a
density matrix but, at the same time, change its correlation properties. As a
consequence the correlation properties can not be decided upon examining the
spectrum of a state. Among density matrices with the same spectra we find
correlated as well as non-correlated states. An answer to a natural question
about the fraction of non-correlated states is the main result of the paper.

To achieve the goal we present a unified scheme for detection of correlations
defined in the above-described way. It has its own merits that will be
elaborated in forthcoming publications, here we describe only the simplest form
of it, suitable for the present purposes.

%

The first step is a proper description of the uncorrelated pure states
$\mathcal{M}$. Observe that there is no observable having vanishing expectation
value only on non-correlated states \cite{badziag02}. Instead, we assume that
the class of non-correlated pure states $\mathcal{M}$ is defined by a condition
involving two copies of a state,
\begin{equation}
\ket{\psi}\ \text{is non-correlated}\ \Longleftrightarrow\bra{\psi}\bra{\psi}A\ket{\psi}\ket{\psi}=0\,,
\label{eq:definition class}
\end{equation}
where $A$ is a suitably chosen projection operator acting in the double tensor
product, $\mathcal{H}\otimes\mathcal{H}$, of the underlying Hilbert space
$\mathcal{H}$. In the following we show that indeed, this is a correct
definition of $\mathcal{M}$ in all considered cases. Our criterion for
detection of correlations in the mixed states takes a particularly simple form:
\begin{equation}
\mathrm{tr}\big(\left(\rho_{1}\otimes\rho_{2}\right)\cdot V\big)>0\ \Longrightarrow
\ \rho_{1}\,\mbox{\text{and} }\rho_{2}
\text{ are correlated},\label{eq:criterion formula}
\end{equation}
where $V=A-\mathbb{P}^{\mathrm{asym}}$  and $\mathbb{P}^{\mathrm{asym}}$
denotes the orthogonal projection onto the two fold antisymmetrization,
$\bigwedge^{2}\mathcal{H}$, of $\mathcal{H}$. A particular choice $\rho_1=\rho_2=\rho$
leads to a quadratic witness of correlations
\begin{equation}
f(\rho)=\mathrm{tr}\big(\left(\rho\otimes\rho\right)\cdot V\big) >0  \Longrightarrow
\ \rho\ \text{ is correlated}.  \label{eq:quadratic witness}
\end{equation}
It is important to note that the criterion is independent on the dimension of
$\mathcal{H}$ and uses only algebraic structure of the set $\mathcal{M}$.

In the following it will be expedient to identify pure states with rank-one
density matrices, i.e., $\ket{\psi}\sim\kb{\psi}{\psi}$ for a normalized
$\ket{\psi}$. Under such na identification the set of uncorrelated pure states
$\mathcal{M}$ can be treated as subset of $\mathcal{H}$, as well as a subset of
the set of rank-one density matrices denoted in the following by
$\mathcal{D}_1(\mathcal{H})$. To keep the notation compact we will alternate
between both interpretations of $\mathcal{M}$, as it usually does not cause
confusion. We will use $\mathcal{D}(\mathcal{H})$  do denote the set of all
states (non-negatively definite, trace-one operators on $\mathcal{H}$).

The set of mixed correlated states can be now identified with the convex hull
$\mathcal{M}^\mathrm{c}$ of $\mathcal{M}$,
\begin{equation}
\mathcal{M}^\mathrm{c}=\left\{\rho=\sum_{i}p_{i}\kb{\psi_i}{\psi_{i}}\,:\,p_i>0,\,
\sum_{i}p_{i}=1,\, \ket{\psi_i}\in\mathcal{M}\right\} \,.
\label{eq:convex hull}
\end{equation}

We are now ready to state two theorems from which we deduce the criterion
\eqref{eq:criterion formula}. We present their proofs in the Supplemental
Material \cite{supp}.
\begin{thm}
\label{thm:Consider-a-class}
Assume that there exists a Hermitian operator acting on
$\mathcal{H}\otimes\mathcal{H}$ such that $\bra v\bra wV\ket v\ket w\le0$ for
all $\ket v\in\mathcal{M}$ and for arbitrary $\ket w\in\mathcal{H}$. Then, for
any state $\rho\in\mathcal{M}^\mathrm{c}$ and for arbitrary non-negatively
defined operator $B$ acting on $\mathcal{H}$, we have
\begin{equation}
\mathrm{tr}\big(\left(\rho\otimes B\right)\cdot V\big)\le0.
\label{eq:basic ineuality}
\end{equation}
\end{thm}
Theorem~\ref{thm:Consider-a-class} gives a straightforward way to construct
linear witnesses of correlations. Indeed, whenever we find a non-negative
operator $B$ for which $\mathrm{tr}\left(\left(\rho\otimes B\right)V\right)>0$
we know that $\rho$ is correlated. Theorem~\ref{thm:Consider-a-class} does not
say anything about the existence of the operator $V$ for a given class of pure
states $\mathcal{M}$. The following theorem ensures that such operator exists
whenever $\mathcal{M}$ is given by the condition (\ref{eq:definition class}).
\begin{thm}
\label{thm: ensurance of existance} Consider the class of pure states
$\mathcal{M}$ defined by the condition \emph{(\ref{eq:definition class})}. The
operator $V=A-\mathbb{P}^{\mathrm{asym}}$ satisfies $\bra v\bra wV\ket v\ket
w\le0$ for all $\ket v\in\mathcal{M}$ and for arbitrary $\ket w\in\mathcal{H}$.
\end{thm}
The above result guarantees that the operator $V=A-\mathbb{P}^{\mathrm{asym}}$
fulfils assumptions of Theorem~\ref{thm:Consider-a-class}. Note that $\tau
V\tau=V$, where $\tau$ is the operator swapping between two factors of the
tensor product $\mathcal{H}\otimes\mathcal{H}$. Using Theorems
\ref{thm:Consider-a-class} and \ref{thm: ensurance of existance} we arrive at
the result given by (\ref{eq:criterion formula}).

Below we give formulas for the operator $A$ for four considered classes of
correlations and accompany them with some exemplary applications for Slater
determinants. For Slater determinants we consider
the depolarisation of an arbitrary pure state of two fermions \cite{Eckert}. 

\textbf{Separable states.} For a system of $L$ distinguishable particles the
Hilbert space is $\mathcal{H}_{d}=\bigotimes_{i=1}^{i=L}\mathcal{H}_{i}$. For
simplicity we assume that all $\mathcal{H}_{i}$ are identical,
$\mathcal{H}_{i}\approx\mathbb{C}^{d}$. Pure separable states are given by
\begin{equation}
\mathcal{M}_{sep}=\left\{ \ket{\phi_{1}}\otimes\ldots\otimes\ket{\psi_{L}}\,|\,\ket{\phi_{i}}\in\mathcal{H}_{i}\right\} \,.\label{eq:sep definition}
\end{equation}
We introduce the notation
\begin{equation}
\mathcal{H}_{d}\otimes\mathcal{H}_{d}=\left(\bigotimes_{i=1}^{i=L}\mathcal{H}_{i}\right)\otimes\left(\bigotimes_{i=1'}^{i=L'}\mathcal{H}_{i}\right)\label{eq:convention}
\end{equation}
where where $L=L'$ and spaces from the second copy of the total space are
labeled with primes. It was proven in \cite{Mintert1} that the set
$\mathcal{M}_{sep}$ is characterized by the condition (\ref{eq:definition
class}) where operator $A$ is given by
\begin{equation}
A_{d}=\mathbb{P}_{d}^{\mathrm{sym}}-\mathbb{P}_{11'}^{+}\mathbb{P}_{22'}^{+}\ldots\mathbb{P}_{LL'}^{+}\,,\label{eq:A dist}
\end{equation}
where $\mathbb{P}_{d}^{\mathrm{sym}}$ projects onto
$\mathrm{Sym}^{2}\left(\mathcal{H}\right)$ and operators
$\mathbb{P}_{ii'}^{+}:\mathcal{H}_{d}\otimes\mathcal{H}_{d}\rightarrow\mathcal{H}_{d}\otimes\mathcal{H}_{d}$
that projects onto the subspace of $\mathcal{H}_{d}\otimes\mathcal{H}_{d}$
completely symmetric under interchange spaces $i$ and $i'$. Applying the above
above result to criterion (\ref{eq:criterion formula}) we recover ``quadratic
entanglement witness'' considered before by, among others, P. Horodecki
\cite{horo true}, F. Mintert, A. Buchleitner \cite{Mintert1}. For a  general
discussion of non-linear entanglement witnesses see also \cite{Agusiak}.
Interesting variation of this method can be found in \cite{badz}.

\textbf{Separable bosonic states }\cite{Eckert}\textbf{.} The relevant Hilbert
space describing the system consisting of $L$ bosonic particles is the $L$-fold
symmetrization of a single-particle $d$-dimensional space,
$\mathcal{H}_{b}=\mathrm{Sym}^{L}\left(\mathbb{C}^{d}\right)$. The set of pure
bosonic separable states are defined by
\begin{equation}
\mathcal{M}_{b}=\left\{ \ket{\phi}\otimes\ket{\phi}\otimes\ldots\otimes\ket{\phi}\,|\,\ket{\phi}\in\mathbb{C}^{d}\right\} \,.\label{eq:bos definition}
\end{equation}
We can treat $\mathcal{H}_{b}$ and $\mathcal{H}_{b}\otimes\mathcal{H}_{b}$
as subspaces of respectively $\mathcal{H}_{d}$ and $\mathcal{H}_{d}\otimes\mathcal{H}_{d}$
defined in the previous part. It was shown \cite{univ frame} that
operator $A$ can be expressed by
\begin{equation}
A_{b}=\mathbb{P}_{b}^{\mathrm{sym}}-\left(\mathbb{P}_{11'}^{+}\circ\mathbb{P}_{22'}^{+}\circ
\ldots\circ\mathbb{P}_{LL'}^{+}\right)\left(\mathbb{P}_{\left\{ 1,\ldots,L\right\} }^{\mathrm{sym}}
\circ\mathbb{P}_{\left\{ 1',\ldots,L'\right\} }^{\mathrm{sym}}\right)\,
\label{eq:bos A}
\end{equation}
where $\mathbb{P}_{b}^{\mathrm{sym}}$ projects onto
$\mathrm{Sym}^{2}\left(\mathcal{H}_{b}\right)$ and $\mathbb{P}_{\left\{
1,\ldots,L\right\} }^{\mathrm{sym}}$ and $\mathbb{P}_{\left\{
1',\ldots,L'\right\} }^{\mathrm{sym}}$ are projectors onto subspaces of
$\mathcal{H}_{d}\otimes\mathcal{H}_{d}$ completely symmetric under interchange
of spaces labeled by indices from the set $\left\{ 1,2,\ldots,L\right\} $ and
$\left\{ 1',2',\ldots,L'\right\} $ respectively.

\textbf{Slater determinants }\cite{Eckert}\textbf{.} The Hilbert space
describing $L$ fermions is the $L$-fold antisymmetrization of the
single-particle $d$-dimensional space,
$\mathcal{H}_{f}=\mathrm{\bigwedge}^{L}\left(\mathbb{C}^{d}\right)$. We
distinguish the class of Slater determinants
\begin{equation}
\mathcal{M}_{f}=\left\{ \ket{\phi_{1}}\wedge\ket{\phi_{2}}\wedge\ldots\wedge\ket{\phi_{L}}|\,
\ket{\phi_{i}}\in\mathbb{C}^{d},\,\bk{\phi_{i}}{\phi_{j}}=\delta_{ij}\right\} \,.
\label{eq:ferm definition}
\end{equation}
As before, we treat $\mathcal{H}_{f}$ and $\mathcal{H}_{f}\otimes\mathcal{H}_{f}$
as subspaces of, respectively, $\mathcal{H}_{d}$ and $\mathcal{H}_{d}\otimes\mathcal{H}_{d}$.
It was proven \cite{univ frame} that in this case the operator $A$
is given by
\begin{align}
&A_{f}=\mathbb{P}_{f}^{\mathrm{sym}}-
\\
&\frac{2^{L}}{L+1}\left(\mathbb{P}_{11'}^{+}\circ\mathbb{P}_{22'}^{+}\circ\ldots
\circ\mathbb{P}_{LL'}^{+}\right)\left(\mathbb{P}_{\left\{
1,\ldots,L\right\} }^{\mathrm{asym}}\circ\mathbb{P}_{\left\{
1',\ldots,L'\right\} }^{a\mathrm{sym}}\right)\,, \nonumber
\label{eq:dermwitness}
\end{align}
where $\mathbb{P}_{f}^{\mathrm{sym}}$ projects onto
$\mathrm{Sym}^{2}\left(\mathcal{H}_{f}\right)$ and $\mathbb{P}_{\left\{
1,\ldots,L\right\} }^{a\mathrm{sym}}$ and $\mathbb{P}_{\left\{
1',\ldots,L'\right\} }^{a\mathrm{sym}}$ are projectors onto subspaces of
$\mathcal{H}_{d}\otimes\mathcal{H}_{d}$ completely asymmetric under interchange
of spaces labeled by indices from the set $\left\{ 1,2,\ldots,L\right\}$ and
$\left\{ 1',2',\ldots,L'\right\}$, respectively. Let us now study arbitrary
depolarized pure states of two fermions. Any state
$\ket{\psi}\in\mathcal{H}_{f}$ can be written \cite{Eckert} as
\begin{equation}
\ket{\psi}=\sum_{i=1}^{i=\left\lfloor \frac{d}{2}\right\rfloor }\lambda_{i}\ket{\phi_{2i-1}}\wedge\ket{\phi_{2i}},\,\label{eq:arbitrary 2 ferm}
\end{equation}
where $\lambda_{i}\ge0$, $\sum_{i=1}^{i=\left\lfloor \frac{d}{2}\right\rfloor
}\lambda_{i}^{2}=1$, and the vectors $\ket{\phi_{i}}$ are pairwise orthogonal.
As an example we consider an arbitrary depolarisation of the state
$\ket{\psi}$,
\begin{equation}
\rho_{\psi}\left(p\right)=\left(1-p\right)\kb{\psi}{\psi}+
p\frac{2\mathbb{I}}{d\left(d-1\right)}\,,\label{eq:slater family}
\end{equation}
where $p\in[0,\,1]$ and $\mathbb{I}$ is the identity operator on
$\mathcal{H}_{f}$. Direct usage of the criterion (\ref{eq:quadratic witness})
shows that the state $\rho_{\psi}\left(p\right)$ is correlated if
\begin{equation}
\left(1-p\right)^{2}\left(5-2\sum_{i=1}^{i=\left\lfloor
\frac{d}{2}\right\rfloor }
\lambda_{i}^{4}\right)+2p(1-p)\chi_{1}(d)+p^{2}\chi_{2}(d)>3,\,
\label{eq:criterion slater depolar}
\end{equation}
where
$\chi_{1}(d)=3+\frac{2\left(d-2\right)\left(d-3\right)}{d\left(d-1\right)}$ and
$\chi_{2}\left(d\right)=\frac{2\left(d+1\right)}{d-1}+\frac{6}{d\left(d-1\right)}$.

\textbf{Fermionic Gaussian states}. Hilbert space describing fermions with
unconstrained number of particles is the Fock space,
$\mathcal{H}_{\mathrm{Fock}}\left(\mathbb{C}^{d}\right)=
\bigoplus_{L=0}^{L=d}\bigwedge^{L}\left(\mathbb{C}^{d}\right)$, where
$\bigwedge^{0}\left(\mathbb{C}^{d}\right)$ is the one dimensional linear
subspace spanned by the Fock vacuum $\ket 0$. Standard fermionic creation and
annihilation operators: $a_{i}^{\dagger}$, $a_{i}$ , $i=1,\ldots,d$, satisfying
canonical anti-commutation relations, act in $\mathcal{H}_{\mathrm{Fock}}$. In
order to define pure fermionic Gaussian states it is convenient to introduce
Majorana fermion operators \cite{ferm lagr,ferm noisy}:
$c_{2k-1}=a_{k}+a_{k}^{\dagger}$, $c_{2k}=i\left(a_{k}-a_{k}^{\dagger}\right)$
, $k=1,\ldots,d$. One checks that they are Hermitian and satisfy the
anticommutation relations $\left\{ c_{k},\, c_{l}\right\} =2\delta_{kl}$. For a
mixed state $\rho$ one defines its correlation matrix $M$,
\begin{equation}
M_{ij}=\frac{i}{2}\mathrm{Tr}\left(\rho\,\left[c_{i},\,
c_{j}\right]\right)\,,\, i,j=1,\ldots,2d\,.\label{eq: correlation matrix}
\end{equation}
The $2d\times 2d$ matrix $M$ is a real and anti-symmetric. Pure fermionic
Gaussian states are, by definition, states for which the correlation matrix is
orthogonal,
\begin{equation}
\mathcal{M}_{g}=\left\{
\ket{\psi}\in\mathcal{H}_{\mathrm{Fock}}\left(\mathbb{C}^{d}\right)\,|\,
MM^{T}=\mathbb{I}_{2d}\right\} ,\,\label{eq:gauss}
\end{equation}
where the matrix $M$ depends on $\ket{\psi}$ via (\ref{eq: correlation matrix})
and $\mathbb{I}_{2d}$ is the identity matrix of size $2d$. Let us define the
Hermitian operator $\Lambda=\sum_{i=1}^{2d}c_{i}\otimes c_{i}$. Let
$\mathbb{P}_{0}$ denote the projector onto the subspace with the eigenvalue
zero of the operator $\Lambda$. From \cite{ferm lagr,ferm noisy} it easily
follows that the operator $A$ has the form
$
A_{g}=\mathbb{P}_{g}^{\mathrm{sym}}-\mathbb{P}_{0}$, 
where $\mathbb{P}_{g}^{\mathrm{sym}}$ projects onto
$\mathrm{Sym}^{2}\left(\mathcal{H}_{\mathrm{Fock}}\right)$. In a recent paper \cite{gaussSim} the set of convex-Gaussian fermionic states was characterized analytically in the first non-trivial case of $d=4$ modes. The method presented here cannot reproduce this result but can be used to detect states that are not convex-Gaussian for arbitrary number of modes.

\begin{figure}[h]
\begin{centering}
\includegraphics[width=6cm]{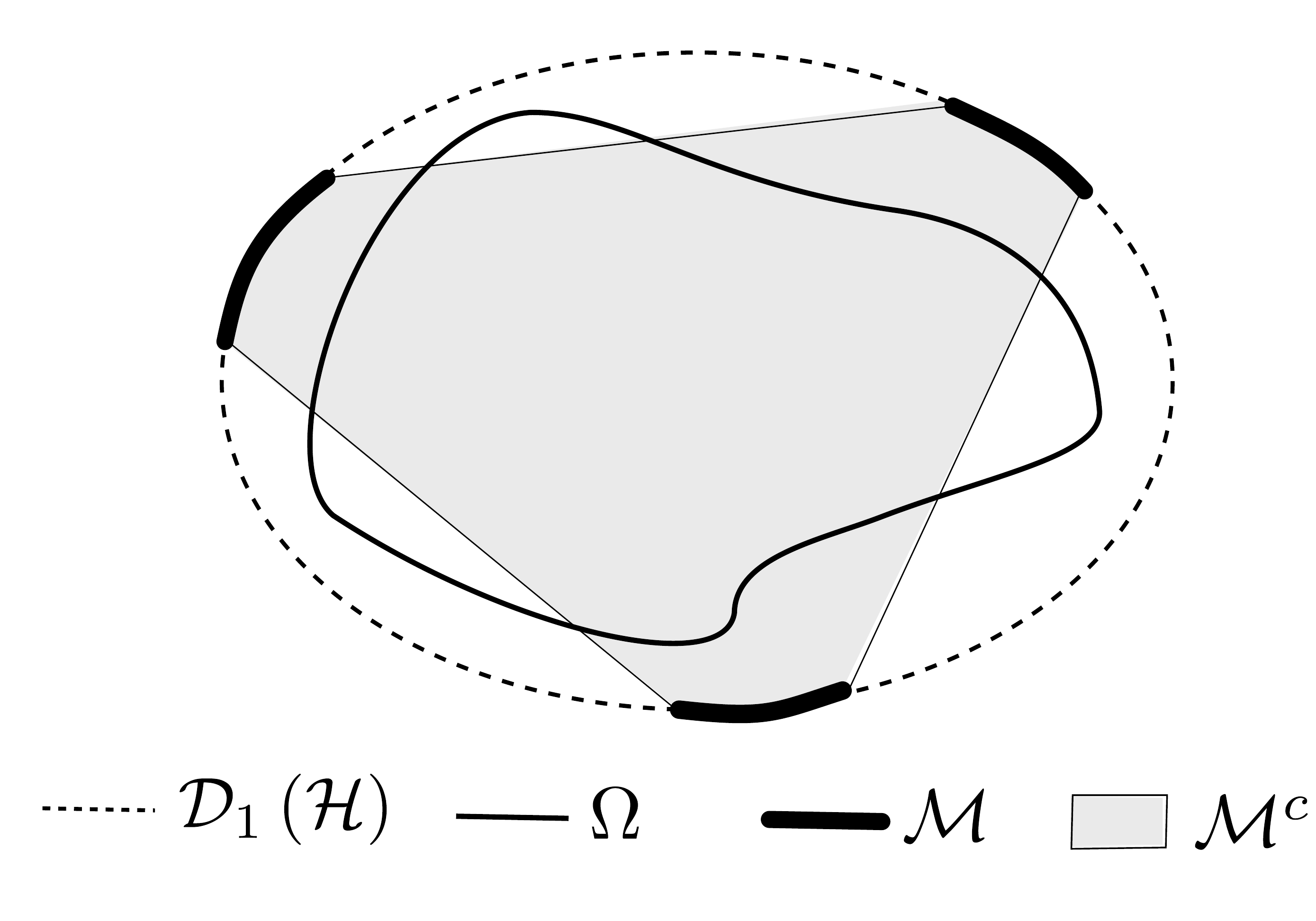}
\caption{\label{fig:Illustration} Illustration of the geometry
of correlated states in the space of density matrices $\mathcal{D\left(\mathcal{H}\right)}$.
The space $\mathcal{D\left(\mathcal{H}\right)}$ is located inside
the region bounded by the dashed line. For simplicity of presentation
we identified the boundary of $\mathcal{D\left(\mathcal{H}\right)}$
with space of pure states $\mathcal{D}_{1}\left(\mathcal{H}\right)$.
The class of pure states $\mathcal{M}$ is given by the thick solid
segments laying on $\mathcal{D}_{1}\left(\mathcal{H}\right)$. The
class of non-correlated states, $\mathcal{M}^{c}$, is marked by a
grey region. Set of isospectral density matrices $\Omega$ is marked by the solid loop.
}

\par\end{centering}

\end{figure}

Having proved the criterion (\ref{eq:criterion formula}) and demonstrating its
usefulness it is natural to ask how often it is satisfied and what does it say
about correlation properties of the system in question. We answer these
questions by studying typical properties of function $f$ (see
(\ref{eq:quadratic witness})) restricted to the set of density matrices having
the same spectrum (isospectral density matrices). We denote by $\Omega_{\left\{
p_{1},\ldots,p_{N}\right\} }$ the set of all density matrices having an
(ordered) spectrum $\left\{ p_{1},\ldots,p_{N}\right\}$, where $p_i$ are real
numbers satisfying $0\le p_{1}\le\ldots\le p_{N}$,~$\sum_{i}p_{i}=1$,
Obviously, we have $\Omega_{\left\{ 1,0,\ldots,0\right\}
}=\mathcal{D}_{1}\left(\mathcal{H}\right)$. On the set $\Omega_{\left\{
p_{1},\ldots,p_{N}\right\} }$ the special unitary group
$SU\mathcal{\left(H\right)}$ acts naturally via the conjugation: $U.\rho=U\rho
U^{\dagger}$. Every two density matrices from $\Omega_{\left\{
p_{1},\ldots,p_{N}\right\} }$ are conjugate in this manner by some element of
$SU\mathcal{\left(H\right)}$. In what follows we will write for short
$\Omega_{\left\{ p_{1},\ldots,p_{N}\right\} } \equiv \Omega $. The geometry of
the considered setting is presented on Figure \ref{fig:Illustration}.

The set $\Omega$ is equipped with a natural unitarily invariant probability
measure $\mu_{\Omega}$ that stems from the (normalized) Haar measure $\mu$ on
$SU\left(\mathcal{H}\right)$ and the transitive action of this group on
$\Omega$. Our strategy is as follows: for each $\Omega$ we employ the
concentration of measure inequality \cite{ledoux,random matrix} for the
function $f_{\Omega}$ which is the restriction of $f$ (see (\ref{eq:quadratic
witness})) to $\Omega$. Having done so we have the information about typical
properties of $f$ on $\Omega$. This gives us, provided the average of
$f_{\Omega}$ is non-negative, the lower bound for the measure of correlated
states on $\Omega$. This insight is different from the previous approaches to
similar problems, usually arising from the entanglement theory, in which
typical properties of the quantity in question (some entanglement measure or
the particular property of a quantum state) were studied on the whole space
$\mathcal{D}\left(\mathcal{H}\right)$ with a particular choice of the
probability measure \cite{vol ent 1,vol ent 2,szarek}. Our reasoning is more
general because it gives the information about typical behaviour of
correlations for each choice of the spectrum. Our final result is the following.

\begin{thm}
\label{thm:estimates} Let the class $\mathcal{M}$ be defined by
(\ref{eq:definition class}) and let $V$ be defined as in Theorem~\ref{thm:
ensurance of existance}. Let $X=\frac{\mathrm{dim}\left(\mathrm{Im}\left(A\right)\right)}{\mathrm{dim}\left(\mathrm{Sym^{2}}\left(\mathcal{H}\right)\right)}$ ,  $P_{\mathrm{cr}}=\frac{1-X}{1+X}$ and let $P\left(\Omega\right)=\sum_{i}p_{i}^{2}$,   denote the purity
of states belonging to $\Omega$. Assume that $P\left(\Omega\right)=P_{\mathrm{cr}}+\delta$ ($\delta>0$).  Then, the following
inequality holds,
\begin{equation}
\mu_{\Omega}\left(\left\{ \rho\in\Omega|\,\rho\ \text{is correlated}\right\} \right)\geq1-\mathrm{exp}\left(-\frac{N\delta^{2}\left(X+1\right)^{2}}{64}\right)\,.\label{eq:final estimate}
\end{equation}
\end{thm}
Here $N$ denotes the dimension of the Hilbert space and $\mathrm{Im}(A)$ the
image of the operator $A$ relevant for the problem in question. The proof of
Theorem \ref{thm:estimates} is presented in the Supplemental Material
\cite{supp}. Values of the relevant parameters appearing in (\ref{eq:final
estimate}) for the four discussed classes of states are presented in Table
\ref{tab:Parameters-characteising-typical}. Value of $X$ for separable states
follows directly from (\ref{eq:A dist}). Value of $X$ for fermionic Gaussian
states \cite{foot} follows easily from the discussion contained in \cite{ferm noisy}. The
origin of the remaining two values is discussed in the Supplemental Material
\cite{supp}. Notice that for separable states, separable bosonic states and
Slater determinants $P_{\mathrm{cr}}\rightarrow 0$ and $N\rightarrow \infty$ as
$L\rightarrow\infty$. To our knowledge results contained in Theorem
\ref{thm:estimates} and Table \ref{tab:Parameters-characteising-typical} were
not obtained elsewhere. Closely related problems in the context of entanglement
theory were considered \cite{vol ent 2} but mostly with the use of numerical
methods.

\begin{table}[h]
\centering{}%
\begin{tabular}{|c|c|c|}
\hline
Class of pure states $\mathcal{M}$ & $N$ & $1-X$\tabularnewline
\hline
Separable states & $d^{L}$ & $2^{1-L}\left(\frac{\left(d+1\right)^{L}}{d^{L}+1}\right)$\tabularnewline
\hline
Separable bosonic states & {  $\binom{d+L-1}{L}$} & {  $1-\frac{2\binom{d+2L-1}{2L}}{\binom{d+L-1}{L}\left(\binom{d+L-1}{L}+1\right)}$ }\tabularnewline
\hline
Slater determinants & $\binom{d}{L}$ & $\frac{2\binom{d}{L}}{\binom{d}{L}+1}\cdot\frac{d+1}{\left(L+1\right)\left(d+1-L\right)}$\tabularnewline
\hline
Fermionic Gaussian states & {$2^{d-1}$} & {$\frac{\binom{2d}{d}}{\left(2^{d-1}+1\right)2^{d-1}}$}\tabularnewline
\hline
\end{tabular}\caption{\label{tab:Parameters-characteising-typical}Parameters characterising
typical behaviour of correlated states that appear in Theorem \ref{thm:estimates}.}
\end{table}

To summarize, we have presented a criterion for detection of correlated mixed
quantum states i.e. states that cannot be expressed as a convex combination of
uncorrelated pure sates belonging to the class $\mathcal{M}$ given by the
operator $A$ (see Eq.(\ref{eq:definition class})). We have demonstrated our
criterion for four physically relevant classes of pure states: separable
states, separable bosonic states, Slater determinants and fermionic Gaussian
states. Moreover, we have shown that our criterion leads to the
characterisation of typical properties of set of correlated states belonging to
the set of isospectral density matrices. Let us end with comments concerning
the obtained results. First, we would like to remark that it is not a
coincidence that the projector $A$ exists in all four considered classes of
pure states. It is a general result in the representation theory of semisimple
Lie groups \cite{lichteinstein,quasiclassical,On detection} that such operator
exists for so-called Perelomov coherent states \cite{Perelomov}, i.e. states
that form the orbit of the relevant symmetry group through the highest weight
vector of a irreducible representation. This observation covers first three
cases as it was discussed in \cite{univ frame}. On the other hand, fermionic
Gaussian states can be also treated as the orbit through the Fock vacuum of the
group of fermionic Bogoliubov transformations or, equivalently, the group
$Pin\left(2d\right)$ \cite{gauss nap}. It is tempting to ask whether the
operator $A$ exists for other physically interesting classes of states (like
Glauber coherent states or bosonic Gaussian states). There are other ways in
which one can generalize the presented approach. For instance, on can try to
subtract in Eq.(\ref{eq:criterion formula}) not $\mathbb{P}^{\mathrm{asym}}$
but some operator that would be more suitable for a given problem. In the
future we plan to extend our framework to cases when $\mathcal{M}$ is given by
the operator acting on many copies of the physical Hilbert space.

We would like to thank Szymon Charzy\'{n}ski, Janek Gutt and Adam Sawicki for
fruitful discussions. The support of the  ERC grant QOLAPS, NCN grant DEC-2013/09/N/ST1/02772 and SFB-TR12
program financed by Deutsche Forschungsgemeinschaft and COST
Action MP 1006 is gratefully acknowledged.

\newpage
\section*{Supplemental Material}

We provide here technical details that were omitted in the main part
of the manuscript. We give proofs of Theorems $1$,
$2$ and $3$. Moreover, we give the
reasoning from which we deduce the forms of
$X=\frac{\mathrm{dim}\left(\mathrm{Im}\left(A\right)\right)}{\mathrm{dim}\left(\mathrm{Sym^{2}}\left(\mathcal{H}\right)\right)}$
in the case of bosonic separable states and Slater determinants.

\subsection*{Proof of Theorem 1}
\begin{proof}
Since the expression (5) is linear in $B$ and every
non-negative operator is of the form $B=\sum_{i}\kb{w_{i}}{w_{i}}$ for some
(not necessary normalized) $\ket{w_{i}}\in\mathcal{H}$, it is enough to
consider $B=\kb ww$, where $\ket w\in\mathcal{H}$ is arbitrary. By definition
the condition $\rho\in\mathcal{M}^{c}$ is equivalent to
$\rho=\sum_{j}p_{j}\kb{v_{j}}{v_{j}}$ for $\ket{v_{i}}\in\mathcal{M}$ and
$p_{j}\ge0$, $\sum_{j}p_{j}=1$. Using that and the assumption (5) about the operator $V$ we get
\[
\mathrm{tr}\left(\left(\rho\otimes\kb ww\right)V\right)=\sum_{j}p_{j}\bra{v_{j}}\bra wV\ket w\ket{v_{j}}\le0\,.
\]
This concludes the proof.
\end{proof}

\subsection*{Proof of Theorem 2}
\begin{proof}
Let $\ket v\in\mathcal{M}$ and let $\ket w\in\mathcal{H}$. Let us
write $\ket w=\ket{v_{||}}+\ket{v_{\perp}}$, where $\ket{v_{||}}\propto\ket v$
and $\ket{v_{\perp}}\perp\ket v$. We have the following equalities
\[
\bra v\bra wA\ket v\ket w=\bra v\bra{v_{||}}A\ket v\ket{v_{||}},
\]
\[
\bra v\bra w\mathbb{P}^{\mathrm{asym}}\ket v\ket w=\bra v\bra{v_{\perp}}\mathbb{P}^{\mathrm{asym}}\ket v\ket{v_{\perp}}=\frac{1}{2}\bk{v_{\perp}}{v_{\perp}}\,.
\]
We have used the fact that operator $A$ is an orthonormal projector
and therefore we have $A\ket v\ket v=A\ket v\ket{v_{||}}=0$. Consequently,
we get the desired inequality

\[
\bra v\bra wV\ket v\ket w=\bra v\bra{v_{\perp}}A\ket v\ket{v_{\perp}}-\frac{1}{2}\bk{v_{\perp}}{v_{\perp}}\le0\,,
\]
where the estimate stems from the fact that $\bra v\bra{v_{\perp}}A\ket v\ket{v_{\perp}}\le\bra v\bra{v_{\perp}}\mathbb{P}^{\mathrm{sym}}\ket v\ket{v_{\perp}}=\frac{1}{2}\bk{v_{\perp}}{v_{\perp}}$,
where $\mathbb{P}^{\mathrm{sym}}$ is the projector onto $\mathrm{Sym}^{2}\left(\mathcal{H}\right)$.
\end{proof}

\subsection*{Proof of Theorem 3}
\begin{proof}
In what follows we use the notation of Theorem 3. By $\mathbb{E}_{\Omega}\left(\cdot\right)$
we denote the expectation value with respect to the probability measure
$\mu_{\Omega}$.
In order to prove (18) we first show that
\begin{equation}
\mathbb{E}_{\Omega}\left(f_{\Omega}\right)=\frac{P\left(\Omega\right)}{2}\left(X+1\right)+\frac{X-1}{2}\, .\label{eq:expectation}
\end{equation}
 We first  construct an auxiliary
function on the unitary group $SU\left(\mathcal{H}\right)$:
\begin{equation}
\tilde{f}_{\Omega}\left(U\right)=f\left(U\rho_{0}, U^{\dagger}\right)\,,\tag{S.1} \label{eq:auxilliary}
\end{equation}
where $\rho_{0}=\mathrm{diag}\left(p_{1},\ldots,p_{N}\right)$ denotes
the diagonal matrix with the spectrum $\left\{ p_{1},\ldots,p_{N}\right\} $.
Let $\mathbb{E}\left(\cdot\right)$ denote the expectation value with
respect to the Haar measure $\mu$. From the definition of the measure
$\mu_{\Omega}$ we have that $\mathbb{E}_{\mu}\left(f_{\Omega}\right)=\mathbb{E}\left(\tilde{f}_{\Omega}\right)$.
Using the definition of the function $f$ and the linearity of the
trace we have
\begin{align}
\mathbb{E}\left(\tilde{f}_{\Omega}\right) & =\int_{SU\left(\mathcal{H}\right)}dU\,\mathrm{tr}\left(U\otimes U\left(\rho_{0}\otimes\rho_{0}\right)U^{\dagger}\otimes U^{\dagger}\, V\right)\nonumber \\
 & =\mathrm{tr}\left(\int_{SU\left(\mathcal{H}\right)}dU\,\left(U\otimes U\left(\rho_{0}\otimes\rho_{0}\right)U^{\dagger}\otimes U^{\dagger}\right)\, V\right)\,. \tag{S.2} \label{eq:intergals long}
\end{align}
The integral inside the trace can be computed via the known \cite{Barut Raczka}
formula valid for an arbitrary operator $X$ acting on $\mathcal{H}\otimes\mathcal{H}$,
\begin{equation}
\int_{SU\left(\mathcal{H}\right)}dU\,\left(U\otimes U\, X\, U^{\dagger}\otimes U^{\dagger}\right)=a\mathbb{P}^{\mathrm{sym}}+b\mathbb{P}^{\mathrm{asym}}\,, \tag{S.3} \label{eq:integral unitary}
\end{equation}
where $a=2\frac{\mathrm{tr}\left(X\,\mathbb{P}^{\mathrm{sym}}\right)}{N\left(N+1\right)}$
and $b=2\frac{\mathrm{tr}\left(X\,\mathbb{P}^{\mathrm{asym}}\right)}{N\left(N-1\right)}$.
Applying (\ref{eq:integral unitary}) to (\ref{eq:intergals long})
and using definitions of operators $V$ and $\rho_{0}$ we obtain
(S.\ref{eq:expectation}).

We now use the inequality
\[
\mu_{\Omega}\left(\left\{ \rho\in\Omega|\,\rho\notin\mathcal{M}^{c}\right\} \right)\geq\mu_{\Omega}\left(\left\{ \rho\in\Omega|\, f_{\Omega}\left(\rho\right)>0\right\} \right)\,,
\]
where $\mu_{\Omega}\left(\left\{ \rho\in\Omega|\, f_{\Omega}\left(\rho\right)>0\right\} \right)$
is the measure of states for which our correlation witness detects
correlations. By definition of the measure $\mu_{\Omega}$, we have
\begin{equation}
\mu_{\Omega}\left(\left\{ \rho\in\Omega|\, f_{\Omega}\left(\rho\right)>0\right\} \right)=\mu\left(\left\{ U\in SU\left(\mathcal{H}\right)|\,\tilde{f}_{\Omega}\left(U\right)>0\right\} \right)\,. \tag{S.4} \label{eq: intermediate estimate}
\end{equation}
If $\mathbb{E}\left(\tilde{f}_{\Omega}\right)$ is positive we can
use a variant of the concentration of measure inequality for the group
$SU\left(\mathcal{H}\right)$ to estimate right hand side of (\ref{eq: intermediate estimate}).
Indeed, for every smooth function $F$ defined on $SU\left(\mathcal{H}\right)$
we have the inequality \cite{ledoux,random matrix}
\begin{equation}
\mu\left(\left\{ U\in SU\left(\mathcal{H}\right)|\, F\left(U\right)>\mathbb{E}\left(F\right)-\epsilon\right\} \right)\ge1-\mathrm{exp}\left(-\frac{N\epsilon^{2}}{4L^{2}}\right)\,, \tag{S.5} \label{eq:concentration of measure}
\end{equation}
where $\epsilon>0$ and $L=\mathrm{max}_{U\in SU\left(\mathcal{H}\right)}\left|\nabla F\right|$.
The gradient is taken with respect to the natural metric induced on
$SU\left(\mathcal{H}\right)$ when it is viewed as a subset of the
algebra of linear operators, $\mathrm{Lin}\left(\mathcal{H}\right)$,
equipped with the Hilbert-Schmidt inner product $\left\langle A,\, B\right\rangle =\mathrm{tr}\left(A^{\dagger}B\right)$.
Note that $\mathbb{E}\left(\tilde{f}_{\Omega}\right)=\delta\left(X+1\right)$
for $P\left(\Omega\right)=P_{\mathrm{cr}}+\delta$. We take $F=\tilde{f}_{\Omega}$
and apply (\ref{eq:concentration of measure}) with $\epsilon=\mathbb{E}\left(\tilde{f}_{\Omega}\right)$,
\begin{equation}
\mu\left(\left\{ U\in SU\left(\mathcal{H}\right)|\,\tilde{f}_{\Omega}\left(U\right)>0\right\} \right)\ge1-\mathrm{exp}\left(-\frac{N\delta^{2}\left(X+1\right)^{2}}{4C^{2}}\right), \tag{S.6} \label{eq:wittness estimate}
\end{equation}
where $C=\mathrm{max}_{U\in SU\left(\mathcal{H}\right)}\left|\nabla\tilde{f}_{\Omega}\right|$.
Inequalities (\ref{eq:wittness estimate}) and (\ref{eq: intermediate estimate})
almost give (18). We complete the proof by observing
that $C\leq4$. Indeed, from definition of the gradient we have:

\begin{equation}
\left\langle \left.\nabla\tilde{f}_{\Omega}\right|_{U},\, YU\right\rangle =\left.\frac{d}{dt}\right|_{t=0}\tilde{f}_{\Omega}\left(U\right)\,, \tag{S.7} \label{eq:gradient 1}
\end{equation}
where $U\in SU\left(\mathcal{H}\right)$, $U(t)=e^{tY}U$ , $Y=-Y^{\dagger}$
and $YU\in T_{U}SU\left(\mathcal{H}\right)$ (the tangent space treated
as a (real) subspace of $\mathrm{Lin}\left(\mathcal{H}\right)$).
It follows that

\begin{equation}
\left\langle \left.\nabla\tilde{f}_{\Omega}\right|_{U},\, YU\right\rangle =\mathrm{tr}\left(\left[Y\otimes\mathbb{I}+\mathbb{I}\otimes Y,\,\rho\otimes\rho\right]\, A\right), \tag{S.8} \label{eq: gradient 2}
\end{equation}
where $\rho=U\rho_{0}U^{\dagger}$ and $\mathbb{I}$ is the identity
operator. When passing from (\ref{eq:gradient 1}) to (\ref{eq: gradient 2})
we have used the fact that $\mathbb{P}^{\mathrm{asym}}=U\otimes U\,\mathbb{P}^{\mathrm{asym}}\, U^{\dagger}\otimes U^{\dagger}$
for an arbitrary unitary operator $U$. Expanding $\rho$ in the eigenbasis
and using the fact that $A$ is an orthonormal projector we get

\begin{equation}
\left|\left\langle \left.\nabla\tilde{f}_{\Omega}\right|_{U},\, YU\right\rangle \right|\leq4\sqrt{\left\langle YU,\, YU\right\rangle }\,. \tag{S.9} \label{eq:almost final}
\end{equation}
Plugging to the above $YU=\left.\nabla\tilde{f}_{\Omega}\right|_{U}$
gets the desired result.
\end{proof}

\subsection*{Values of $X$ for bosonic separable states and Slater determinants}

We give here the reasoning justifying formulas for
$X=\frac{\mathrm{dim}\left(Im\left(A\right)\right)}{\mathrm{dim}\left(\mathrm{Sym}^{2}\left(\mathcal{H}\right)\right)}$
appearing in (18). We use the fact that the operator $A$
has a clear group theoretical interpretation in these two cases. One can
identify $\mathcal{H}_{b}$ and $\mathcal{H}_{f}$ with the carrier spaces of
irreducible representations of $SU\left(d\right)$ \cite{quasiclassical,univ
frame}. All irreducible representations of $SU\left(d\right)$ are parameterized
by so-called highest weights \cite{Barut Raczka}, i.e. non increasingly ordered
sequences of length $d-1$ consisting of non-negative integers:

\begin{equation}
\lambda=\left(\lambda_{1},\lambda_{2},\ldots,\lambda_{d-1}\right)\,, \tag{S.10} \label{eq:highest weight}
\end{equation}
where $\lambda_{1}\ge\lambda_{2}\ge\ldots\ge\lambda_{d-1}\,,$ and
$\lambda_{i}\in\left\{ 0,\,1,\,\ldots\right\} $. One can also represent
$\lambda$ by a Young diagram \cite{Young tableaux} - a collection
of boxes arranged in left-justified rows, with non increasing lengths
when looked  from the top to the bottom. In what follows we will use
the notation $2\lambda=\left(2\lambda_{1},2\lambda_{2},\ldots,2\lambda_{d-1}\right)$.
Clearly, $2\lambda$ is also a highest weight so it corresponds to
somme irreducible representation of $SU\left(d\right)$.

The highest weights corresponding to $\mathcal{H}=\mathcal{H}_{b}$
and $\mathcal{H}=\mathcal{H}_{f}$ describing respectively $L$ bosons
and $L$ fermions are
\begin{equation}
\lambda_{b}=\left(L,0,\ldots,0\right)\,\text{and}\,\lambda_{f}=\left(1,\ldots1,0,\ldots0\right)\,, \tag{S.11} \label{eq:weights}
\end{equation}
where there are precisely $L$ ones appearing in the formula for $\lambda_{f}$.
In \cite{On detection} it was proven that the operator
$A\in\mathrm{Herm}\left(\mathrm{Sym}^{2}\left(\mathcal{H}\right)\right)$
defining families of separable bosonic states and Slater determinants is given
by

\begin{equation}
A=\mathbb{P}^{\mathrm{sym}}-\mathbb{P}^{2\lambda},\,\label{eq:structure A} \tag{S.12} 
\end{equation}
where $\mathbb{P}^{2\lambda}$ is the projector onto the unique irreducible
representation of the type $2\lambda$ that appear in $\mathrm{Sym}^{2}\left(\mathcal{H}\right)$
(treated as a carrier space of a representation of $SU\left(d\right)$).
Therefore, the problem of computing $X$ for bosonic separable states
and Slater determinants reduces to computation of dimensions of representations
of $SU\left(d\right)$ described by highest weights $2\lambda_{b}$
and $2\lambda_{f}$ respectively. We perform these computations explicitly
with the usage of methods described in \cite{Cvitanovic}.






\end{thebibliography}


\begin{thebibliography}{References}
\bibitem{foot}    {From \cite{ferm noisy} it follows that all convex-Gaussian states in $\mathrm{Fock}\left(\mathbb{C}^d\right)$ have a block structure with respect to the decomposition $\mathrm{Fock}\left(\mathbb{C}^d\right)=\mathrm{Fock}_+\left(\mathbb{C}^d\right)\oplus \mathrm{Fock}_-\left(\mathbb{C}^d\right)$, where $\mathrm{Fock}_\pm\left(\mathbb{C}^d\right)$ denote the subspaces of the Fock space spanned  by even and respectively odd number of excitations. For this reason fraction of convex-Gaussian states in $\Omega$ equals $0$. Therefore, in order not to consider a trivial situation we consider states and 'global unitary operations' defined solely on $\mathrm{Fock}_+\left(\mathbb{C}^d\right)$.  }
\end{thebibliography}
\end{document}